# Estudi de la Dependència Energètica de les Diferències entre Jets de Quarks i Gluons Utilitzant el detector DELPHI de LEP

Phd. Thesis presented by S. Martí i García[1])

**Abstract**

Three jet events arising from decays of the $Z^0$ boson, collected by the DELPHI detector at LEP, were used to measure differences in the properties of quark and gluon jet fragmentation. Gluon jets were anti-tagged in $b\bar{b}g$ events, by identifiying $b$ quark jets with high purities. Unbiased quark jets came from events $q\bar{q}\gamma$ with two jets plus one photon. A comparison of quark and gluon jet properties in different energy ranges was performed for the first time and within the same detector. The average value of the ratio of the mean charged multiplicities of gluon and quark jets was measured to be

$$1.232 \pm 0.026(\text{esta.}) \pm 0.018(\text{sist.})$$

where the fraction of b-quark initiates jets was 11% and the Durham jet finding algorithm has been used for the selection of three jet events. In agreement with QCD an increase of this ratio with energy was observed at a $3\sigma$ level. A further dependence of this ratio related with the angular acceptance of the algorithm used to reconstruct jets was also measured.

Gluon jets have a broader energy and particle flow around its jet axis than quark jets of equivalent energy.

The string effect has been observed in fully symmetric three-jet events. The ratio $R_\gamma$ of the charged particles flow in the $q\bar{q}$ inter-jet region of the $q\bar{q}g$ and $q\bar{q}\gamma$ samples was measured in agreement with the perturbative QCD expectation

$$\frac{(N_{qq})_{q\bar{q}g}}{(N_{qq})_{q\bar{q}\gamma}} = 0.058 \pm 0.06(\text{stat.+sist.})$$

The dependence of the mean charged multiplicity on the hadronic center-of-mass energy was analysed in photon plus $n$-jet events. A value for $\alpha_s(M_Z)$ has ben determined from these data using a QCD prediction wich includes corrections at leading an next-to-leading order.

$$\alpha_s(M_Z) = 0.116 \pm 0.003(\text{stat.}) \pm 0.009(\text{sist.})$$

---

[1] e-mail: martis@evalo1.ific.uv.es